\documentclass[prl, twocolumn, showpacs, amsmath]{revtex4}
 \pdfoutput=1
\usepackage{dcolumn}
\usepackage{bm}
\usepackage{graphicx}
\usepackage{color}

\DeclareGraphicsExtensions{. jpg,. pdf, . mps, . png, . eps, . ps, . EPS}

\begin{document}
\def\be{\begin{equation}}
\def\ee{\end{equation}}

\def\bc{\begin{center}} 
\def\ec{\end{center}}
\def\bea{\begin{eqnarray}}
\def\eea{\end{eqnarray}}
\newcommand{\avg}[1]{\langle{#1}\rangle}
\newcommand{\Avg}[1]{\left\langle{#1}\right\rangle}

\title{On the criticality of inferred models}

\author{Iacopo Mastromatteo$^1$ and Matteo Marsili$^2$}

\affiliation{$^1$ International School for Advanced Studies, via Beirut
 2/4, 34014, Trieste, Italy\\
$^2$ The Abdus Salam International Center for Theoretical 
Physics, Strada Costiera 11, 34014 Trieste, Italy}

\begin{abstract}
Advanced inference techniques allow one to reconstruct the pattern of interaction from high dimensional data sets, which probe simultaneously thousands of units of extended systems -- such as cells, neural tissues or financial markets.
We focus here on the statistical properties of inferred models and argue that inference procedures are likely to yield models which are close to singular values of parameters, akin to critical points in physics where phase transitions occur. 
These are points where the response of physical systems to external perturbations, as measured by the susceptibility, is very large and diverges in the limit of infinite size. We show that the reparameterization invariant metrics in the space of probability distributions of these models (the Fisher Information) is directly related to the susceptibility of the inferred model. As a result, distinguishable models tend to accumulate close to critical points, where the susceptibility diverges in infinite systems. This region is the one where the estimate of inferred parameters is most stable. In order to illustrate these points, we discuss  inference of interacting point processes with application to financial data and show that sensible choices of observation time-scales naturally yield models which are close to criticality.
%
\end{abstract}

\maketitle

The behavior of complex systems such as a cell, the brain or a financial market, is the result of the pattern of interaction taking place among its components. Technological advances, either in experimental techniques or in data storage and acquisition, have made the micro scale at which the interaction takes place accessible to empirical analysis. Massive data, probing for instance the expression of genes in a cell \cite{Braunstein2008}, the structure and the interactions of proteins \cite{Socolich2005}, the activity of neurons in a neural tissue \cite{Schneidman2006,Cocco2009} or the one of traders in financial markets \cite{Lillo2008}  is now available. This, in principle, makes the reconstruction of the network of interactions at the micro scale possible. 
The reconstruction consists in {\em inferring} a model, specifying the wiring of the network of interactions between micro-units, as well as their strength. 

The typical situation is one where the micro-state is specified by a vector $\vec s$, with component $s_i$ specifying the state of unit $i$, and data consists of a sequence $\hat s=\{\vec s^{(t)},~t=1,\ldots,T\}$ of $T$ samples. Under the assumption that samples can be considered as independent, the problem consists in estimating the probability distribution of $\vec s$, in a way which allows for robust generalization, i.e. for the generation of yet unseen samples.

As a mean of illustration, it is instructive to discuss a specific example. Prices of stocks in a financial market move in a correlated fashion. This correlation arises from the correlated activity of traders buying and selling the different stocks. So for example, a particular activity pattern on stock $1$ may be interpreted as revealing some information to traders, which may induce them to trade stock $2$. One way to formalize this idea in a statistical model, is to fix a time interval $\Delta t$ and define a binary variable on each stock which takes value $+1$ if a trade occurred in that interval, or $-1$ otherwise. In this way, the activity of a stock market with $N$ stocks is represented as a string of $N$ ``spins'' $s_i=\pm 1$, and repeated measurements produce $T$ samples of $\vec s$. 

In spite of its abundance, data is far from being able to completely identify the correct model, and one is left with a complex inference problem. This is because the number of available samples is way smaller than the number of possible microscopic states. In our workhorse example, even reducing attention to the $N=100$ most traded stocks, and at very high frequency $\Delta t=30 s$, one year of data amounts to $T\approx 10^5$ samples, whereas the number of possible micro-states is $2^{100}\approx 10^{30}$. 

This problem is addressed in statistical learning theory in two steps: {\em i)} model selection and {\em ii)} inference of parameters. Boltzmann learning \cite{Ackley1985} addresses {\em i)} by first identifying those empirical quantities which we want the model to reproduce and then invoking the principle of maximum entropy \cite{Kappen1998}. So, for example, if correlated activity on stocks is the result of interaction mediated by traders, and we assume that traders react to movement on single stocks, it is natural to require that our model reproduces the observed pairwise correlations between stocks. If we require that the distribution of $\vec s$ reproduces the measured values of a collection $\Phi=\{\phi^\mu(\vec s)\}_{\mu=1}^M$ of $M$ functions of the micro-state $\vec s$, then maximal entropy prescribes distributions of the exponential form:
\begin{equation}
\label{Prob}
p(\vec s|\underline{g}) = \exp \left( \sum_{\mu=0}^{M} g^\mu \,\phi^\mu (\vec s) \right),
\end{equation}
where $\underline{g}=\{g^\mu: \mu=0\ldots M\}\in\mathcal{G}$ are the parameters of the model, to be inferred in {\em ii)} (see below).

Our focus here, is not on model selection nor on the inference procedure. Rather, we focus on the statistical properties which we expect to observe in inferred models, and argue that there are reasons to expect them to be very peculiar.

Probability distribution of the form (\ref{Prob}), in the limit $N\to\infty$, have been the object of enquiry in statistical mechanics, since its very beginning, in particular, as a function of ``temperature'' which in physics modulates the strength of the interactions between variables. A fictitious (inverse) temperature can be introduced with the replacement $\underline g\to\beta\underline g$. Then $p(\vec s|\beta\underline{g})$ is expected to interpolate between a ``low temperature'' behavior ($\beta\to\infty$), where the distribution is concentrated on few states and the $s_i$ are strongly correlated, and a ``high temperature'' behavior ($\beta\to 0$), where the different components of $\vec s$ are very weakly dependent. These two polar behaviors, often, do not morph continuously into each other as $\beta$ varies in $[0,\infty)$ but rather they do so in a sharp manner, in a small neighborhood of a critical inverse temperature $\beta_c$. The {\em critical} point $\beta_c$ is characterized by the fact that fluctuations -- corresponding e.g. to specific heat in physics -- become very large.

\begin{figure}[h] 
   \centering
   \includegraphics[width=3in]{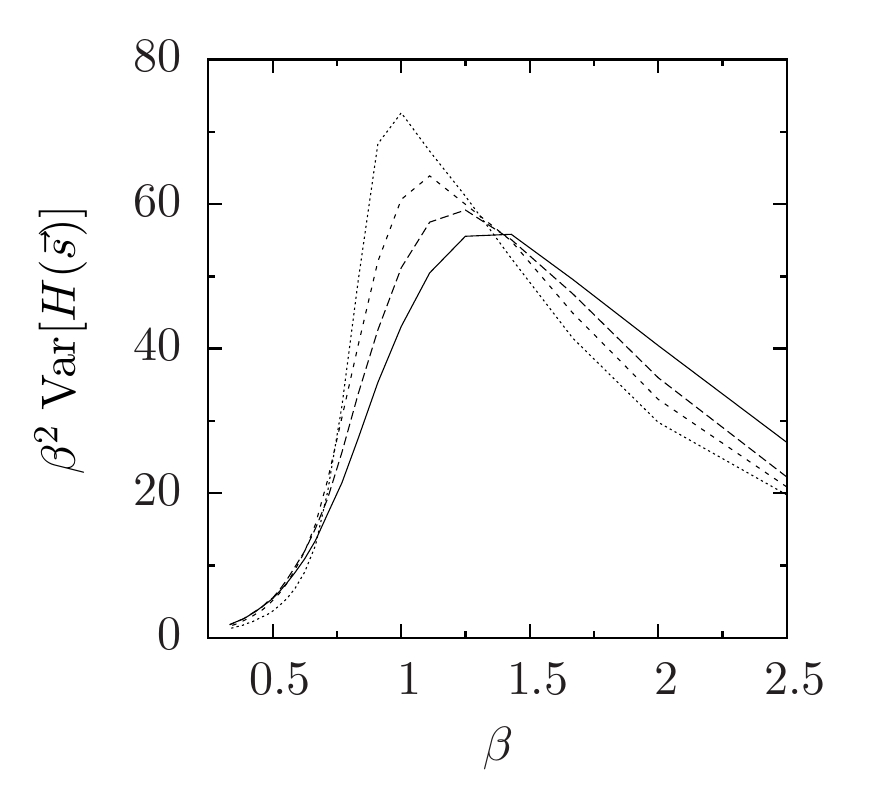} 
   \caption{Specific heat as a function of the inverse temperature $\beta$ for financial data, for various choices of the bin sizes (lines from the top to the bottom correspond
   respectively to $\Delta t = 30, 28, 26,24$ s).}
   \label{fig:TD}
\end{figure}

Remarkably, inference procedure often produces models which seem to be ``poised close to a critical point'', i.e. for which fluctuations are maximal for $\beta\approx 1$, suggesting with $\beta_c\approx 1$. This was first observed in \cite{Tkacik2006} for the activity of neuronal tissues and in \cite{Stephens2008} for the statics of natural images.
Fig. \ref{fig:TD} presents similar evidence for the activity pattern of 100 stocks of the New York Stock Exchange at high frequency (see caption and discussion below for details).

Critical models of the form (\ref{Prob}) seem to be rather special in physics, since they arise only when the parameters are fine tuned to a set of zero measure. In spite of this, they have attracted and still attract considerable interest, with much efforts being devoted to elucidate their properties. On the contrary, critical models seem to arise ubiquitously in the analysis of complex systems \cite{Mora2010}, evoking theories of Self-Organized Criticality \cite{Bak1987}.

Leaving aside the mechanisms for which a real system self-organizes close to a critical point, we address here the issue from a purely statistical point of view. So, for example, when can one conclude in a statistically significant manner that a given system is critical? And then, can criticality be induced by the inference procedure? 

Specifically,
drawing from results of information geometry \cite{Amari1985,Balasubramanian1997} we argue that when the distance in the space of models is
properly defined in a reparametrization invariant manner, one finds that the number of statistically distinguishable models accumulates close to the region in parameter space where models are critical. 
Conversely, models far from the critical points can hardly be distinguished. Loosely speaking, models that can be inferred are only in a finite region around critical points. This implies that when the distance from the critical point is measured in terms of distinguishable models, inferred models turn out to be typically much further away from criticality than what the distance of estimated parameters from criticality would suggest. 
This provides an alternative characterization of criticality, whose relation with information theory was earlier investigated in \cite{Ruppeiner1995} in the context of
thermodynamic fluctuation theory, and in \cite{Zanardi2007} in relation to quantum phase transitions. Indeed our discussion just relies on basic properties of statistical mechanics and
large deviation theory, and doesn't require any  specific assumption about the model it is applied to.

In what follows, we address the problem of the inference of a probability distribution over a set of binary variables. We  recall \cite{Balasubramanian1997} that the statistical distinguishability of empirical distributions, naturally leads to the notion of curvature in the space of probability distributions. Then we show that curvature is related to susceptibility of the corresponding models. 
We apply these ideas to the model of a fully connected ferromagnet, which despite being simple enough to provide tractable solution, realizes all the features previously described. We illustrate the points above by specializing to inference of data produced by ``fully connected'' Hawkes point-processes \cite{Hawkes1971a}.
This shows that when a fictitious temperature is introduced in estimated models, a maximum of the specific heat of the corresponding Ising ferromagnet naturally arises for $\beta \approx 1$.

\section{Distinguishability of statistical models}
Given a set $\hat s=\{\vec s^{(t)}, t=1,\ldots,T\}$ of $T$ observations of  a string of $N$ binary variables $\vec s\in \{ -1,1\}^N$, we consider the problem of estimating from empirical data a statistical model $\mathcal{M}=\{\Phi,\mathcal{G}\}$ defined by a probability distribution as in Eq. (\ref{Prob})
where 
$\Phi=\{\phi^\mu(\vec s)\}_{\mu = 0}^M$  is a collection of functions of the vector $\vec s \in \{ -1,1\}^N$ and $\underline{g}=\{g^\mu: \mu=0\ldots M\}\in\mathcal{G}$ are the corresponding parameters. With the choice $\phi^0(\vec s)=1$, the normalization condition fixes $g^0(\underline g)$ to be the free energy of the Hamiltonian $H(\vec s)=-\sum_{\mu>0}g^\mu \,\phi^\mu(\vec s)$ at temperature equal to one. We shall denote by $\langle\ldots\rangle_{\underline g}$ averages taken over the distribution $p(\vec s|\underline{g})$.

We briefly recall the arguments of Ref. \cite{Balasubramanian1997} to assess whether $\underline{g}$ and $\underline{g}'$ are distinguishable. 
Imagine that the inference procedure returns $\underline{g}$ as the optimal parameters of the distribution and consider resampling a set $\hat{s}_g$ of $T$ observations from $\underline{g}$. 
By Sanov's theorem \cite{Mezard2009}, the probability that the empirical distribution of the sample $\hat{s}_g$ generated by $T$ i.i.d. draws from $p(\vec s|\underline{g})$ falls in a close neighborhood of $\underline{g}'$ is given in the large $T$ limit by 
$p(\hat s_g|\underline{g}')\sim e^{-TD(\underline{g}||\underline{g}')}$, where $D(\underline{g}||\underline{g}')=\left\langle \log\frac{p(\vec s|\underline{g})}{p(\vec s|\underline{g}')}\right\rangle_{\underline g}$ 
is the Kullback-Leibler distance of the two distributions. Requiring that this probability be less than a threshold $\epsilon$, for $\underline{g}$ and $\underline{g}'$ to be distinguishable, implies $D(\underline{g}||\underline{g}')\le \kappa/T$ with $\kappa=-\log\epsilon$. 
This condition identifies a volume of parameters around $\underline{g}$ of distributions which cannot be distinguished from $\underline{g}$, for a finite data set. Since we assumed
$T\gg 1$, this volume can be computed from the expansion 
\[
D(\underline{g}||\underline{g}+\underline\eta)=\frac{1}{2}\sum_{\mu,\nu>0} \eta^\mu \chi^{\mu,\nu}\eta^\nu+O(\eta^3)
\]
where $\hat\chi$ is the matrix of second derivatives of $D(\underline{g}||\underline{g}+\underline\eta)$ computed in $\underline\eta=0$, and is known as the Fisher Information (FI). The volume of distributions which are undistinguishable from $\underline g$ is given by \cite{Balasubramanian1997}:
\begin{equation}
\label{Vol}
\Delta V_{T,k}=\left(\frac{2\pi \kappa}{T}\right)^{M/2}\frac{1}{\Gamma(M/2+1)\sqrt{{\rm det}\hat\chi}}.
\end{equation}
In the language of statistical mechanics, FI corresponds to a generalized susceptibility, and via fluctuation-dissipation relations, to the covariance of operators $\phi^\mu(s)$:
\begin{eqnarray*}
\chi^{\mu,\nu} &=&  - \frac{\partial^2 g^0}{\partial g^\mu \partial g^\nu} =\frac{\partial \langle\phi^\mu(s)\rangle_{\underline g}}{\partial g^\nu}\\
&=& \langle (\phi^\mu (\vec s ) - \langle \phi^\mu \rangle_{\underline g} ) (\phi^\nu  (\vec s ) - \langle \phi^\nu \rangle_{\underline g}) \rangle_{\underline g}.
\end{eqnarray*}
Models with a large FI correspond to models with high susceptibility for which the error on the estimated couplings is small. More precisely, the Cram\'er-Rao bound \cite{Mezard2009} states that given a set of
$T$ independent observation and an unbiased estimator of the couplings $g^*$, its covariance matrix satisfies ${\rm Cov} (g^*) \geq {\hat\chi^{-1}}/{T}$ where, the notation $\hat A \geq \hat B$ indicates that the matrix $\hat A-\hat B$ is positive semidefinite.

Summarizing, the FI provides a parameterization invariant metrics in the space of statistically distinguishable distributions $d\tau = \prod_\mu dg^\mu \sqrt{\det \hat \chi(\underline g)}$. This measure concentrates around the ``critical'' points $\underline g$ where the susceptibility is large (or diverges for $N\to\infty$), which correspond to points where estimates of parameters are more precise. On the contrary, since the susceptibility decreases fast away from critical points, the volume element $d\tau$ is expected to be non-negligible only in a bound region of space. The outcome of the inference procedure can
be considered meaningful when the susceptibility is sufficiently large, or equivalently, when the error in the inferred coupling is small enough. This suggests a maximum distance from the critical point at
which parameters can be inferred \cite{MDLfoot}. 

%
%
%

\subsection{The case of fully connected spin models}

In order to illustrate these concepts, let us consider the simple case of a fully connected ferromagnet 
characterized by the operators $\Phi = \{ 1,\frac{1}{N} \sum_{i<j} s_i s_j , \sum_i s_i \}$ and the corresponding couplings $\underline g = \{-\log Z, J , h \}$. The calculation of the FI is straightforward and, for $N\gg 1$, to leading order, one finds that the invariant measure of distinguishable distributions is given by
\[
d\tau = \sqrt{N/2}\left[\Gamma^{3/2} +A(J)\theta(J-1)\delta(h)\right]dJ dh
\]
where $\Gamma=\frac{\partial m}{\partial h}=\frac{1-{m}^2}{1-J(1-{m}^2)}$ is the spin susceptibility and $m(J,h)=\tanh[Jm(J,h)+h]$. The $\delta(h)$ contribution arises from the discontinuity of $m$ at $h=0$ in the ferromagnetic region $J>1$ with $A(J)= \sqrt{2 \pi^2 m^2 \Gamma}$.

In the highly magnetized region ($J \gg 1$, $h \not = 0$),
the non-singular contribution to the density of states $d\tau \approx \sqrt{8 N} e^{-3 (J+|h|)} dJ dh$ can be explicitly integrated to obtain the number of distinguishable states in a finite region of the phase space. For example,  the number of distinguishable states in the semiplane $J\ge J_{\max}$ stripped of the $h \approx 0$ line, given $T$ observations and a threshold $\kappa$ is $D\approx T \sqrt{N} \; \frac{\sqrt{8}}{9 \pi \kappa} e^{- 3 J_{max}}$. This means that it is not possible to meaningfully infer any value of $J$ greater than $J_{\max} \sim \frac{1}{3} \log \left(T\sqrt{N}\right)$. Under the hypothesis that $h = 0$, instead, we find that $J_{\max}\sim \log \left(T\sqrt{N}\right)$. The volume element $d\tau$ diverges in a non-integrable manner close to the critical point $(J,h) = (1,0)$. For $h=0$ we find $d\tau \sim |1-J |^{-3/2}$ while approaching the critical point on the line $J=1$, one finds the milder divergence $d\tau \sim |h|^{-1}$. 
Hence, there is a macroscopic number of models located in an infinitesimal region around the critical point $(J,h)= (1,0)$. The singularity is smeared by finite size effects when $N<+\infty$, but it retains the main characteristics discussed above. A plot of the density of models for $N=100$ is shown in Fig. \ref{fig:ResultCP}. 

\section{Application to Hawkes processes and real data}

In order to investigate the implications of this picture on the inference of models from data we address the specific case of synthetic data generated by Hawkes processes \cite{Hawkes1971a}. An $N$ dimensional Hawkes process is a generalized Poisson process where the probability of events ${\rm P}\{ d N^i_t = 1| N_t \} = \lambda^i_t \, dt$ in an infinitesimal time interval $dt$ depends on a rate $\lambda^i_t$ which is itself a stochastic variable
\[
\lambda_t^i = \mu^i + \sum_j \int_{-\infty}^t dN^j_{u} \, K^{ij}_{t-u}
\]
which depends on the past realization of the process (here $\mu^i \geq  0$, $K^{ij}_u \geq 0$). This process reproduces a cross-excitatory interaction among the different channels, akin to that occurring between stocks in a financial market \cite{Bowsher2002} or neurons in the brain \cite{Rieke1997}. For our purposes, it will serve as a very simple toy model to generate data with controlled statistical properties, of a similar type to that collected in more complex situations. In fact, the linearity of the model makes it possible to derive analytically some properties in the stationary state. We focus on a fully connected version of the Hawkes process, with $\mu^i = \mu$ and $K^{ij}_u = \frac{\alpha}{N} e^{-\nu u}$. The expected number of events per unit time is $\langle \lambda^i_t\rangle = \frac{\mu}{1-\alpha/\nu}$, and it diverges for $\alpha \rightarrow \nu$. 
We remark that this singularity is not a proper phase transition, as it occurs for any finite $N$. 

We also estimate the activity pattern of an ensemble of 100 stocks of NYSE market (see \cite{Borghesi2005} for details on the dataset). We consider the jump process defined by the function $N^i_t$ which counts  the number of times in which stock $i$ is traded in the time interval $[0,t]$, disregarding the (buy or sell) direction of the trade. Data refers to 100 trading days (from 02.01.2003 to 05.30.2003), of which only the $10^4$ seconds of the central part of the day were retained, in order to avoid the non-stationary effects due to the opening and closing hours of the financial market (see \cite{Bowsher2002}). 

Following  \cite{Schneidman2006}, we map a data-set of events into a sequence of spin configurations, by fixing a time interval $\Delta t$ and setting $s^i_t=+1$ if $\Delta N^i_t=N^i_{t+\Delta t}-N^i_t>0$ and $s^i_t=-1$ if no event on channel $i$ occurred in $[t,t+\Delta t)$. 
The choice of $\Delta t$ fixes not only the number of data points $T= U / \Delta t$, where $U=10^6$ seconds is the total length of the time series. It also fixes the scale at which
the dynamics of the system is observed: for $\Delta t \rightarrow 0$ the system is non-interacting, and can be successfully described by an independent model \cite{Roudi2009a},
while after a certain time scale correlations start to emerge \cite{Epps1979}. Indeed the product of the bin size $\Delta T$ with the event rate $\lambda$ also controls the average magnetization of the system, which can accordingly be tuned from $-1$ to $1$. Hence, as $\Delta t$ varies, the inferred model performs a trajectory in the space of couplings. 

We fit both data with a model of pairwise interacting Ising spins, with operators 
$\Phi = \{1\} \cup \{ s_i \}_{i=1}^N \cup  \{s_i s_j \}_{i<j =1}^N$
and the corresponding couplings 
$\underline g = \{-\log Z \} \cup \{h_i\}_{i=1}^N \cup \{J_{ij} \}_{i<j=1}^N$. Several approximate schemes have been proposed to compute efficiently the maximum entropy estimate of the couplings \cite{Ackley1985}. Here we resort to mean-field theory, which turn out to give results which are consistent with more sophisticated schemes.

\begin{figure}[h] 
   \centering
   \includegraphics[width=3in]{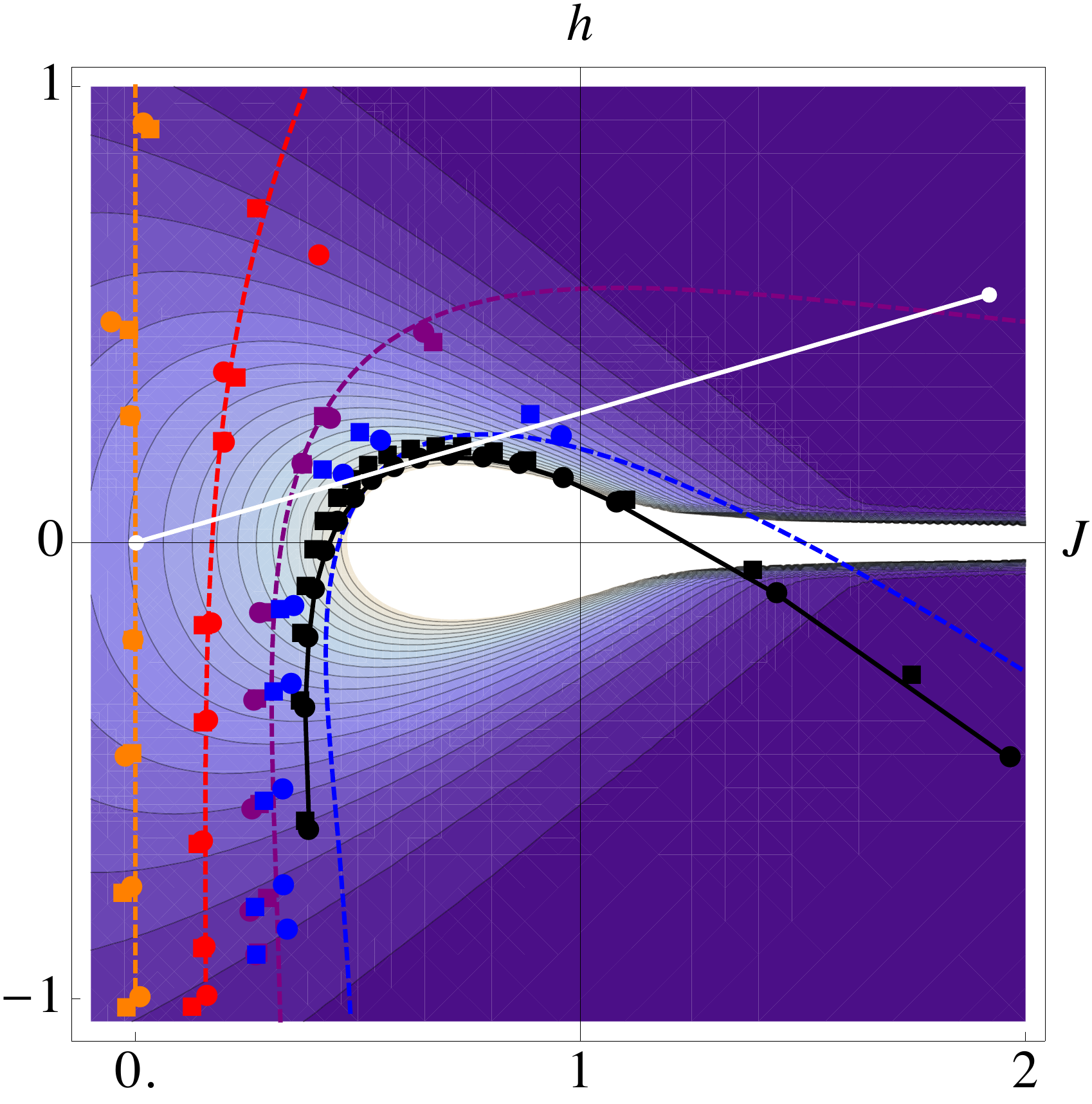} 
   \caption{Mean couplings $(\bar J, \bar h)$ produced by the inference procedure in various cases and with various bin sizes. Orange, red, purple and blue points correspond to the inferred values
   for a simulated Hawkes process with $\nu = 0.3$, $\mu = 0.1$ and $\alpha$ respectively equal to (0, 0.075,0.15,0.225). Boxes and circles correspond respectively to the to the fit of a fully connected model 
   (2  parameters, $\bar J$ and $\bar h$) and to the mean couplings for a spin glass ($N$ fields $h_i$ and $N(N-1)/2$ couplings $J_{ij}$).
   In each of those process we considered $N=100$ channels producing $5000$ events
   each. The dashed line correspond to theoretical, approximate predictions for the inferred couplings of those processes at $T=\infty$. The black points correspond to the values obtained for $U = 10^6 $ seconds of
   financial data corresponding the activity pattern of 100 stocks of the NYSE. On the background, the density of models for a fully connected model is also plotted for the sake of comparison. The white line intersects
   the origin and the inferred values of $(\bar J,\bar h)$ at $\Delta T = 18$ s: for such a choice of the bin size, a fully connected model would have the maximum density of models exactly at $\beta = 1$.}
   \label{fig:ResultCP}
\end{figure}

Fig. \ref{fig:ResultCP} reports the results of the inference on simulated Hawkes processes and of financial data (see caption for details). Given the inferred parameters, we compute the average couplings
 $\bar J = \frac{2}{N(N-1)} \sum_{i<j} J_{ij}$, $\bar h = \frac{1}{N} \sum_i h_i$ and report the trajectory of the point $(\bar J, \bar h)$ as $\Delta t$ varies, for both cases. Fitting a fully connected model $\Phi = \{ 1,\frac{1}{N} \sum_{i<j} s_i s_j , \sum_i s_i \}$, $\underline g = \{ -\log Z, \bar J, \bar h\}$ produces essentially the same results \cite{HWKfoot}.

The region of $\Delta t$ where non-trivial correlations are present but where the binary representation of events is still meaningful \cite{Roudi2009a} corresponds to the region where the trajectory $(\bar J,\bar h)$ is closest to the critical point of a fully connected ferromagnet $(J,h) = (1,0)$. By Cram\'er - Rao's bound, this is also the region of $\Delta t$ where the inferred couplings are likely subject to the smallest errors.
For Hawkes processes, a time interval $\Delta t$ smaller than $1/\nu$ does not allow the process to develop correlations, and for intervals $\Delta t\gg1/\langle\lambda^i\rangle$ the binary nature of the process is lost \cite{HWKfoot}.
In addition, 
{\em i)} increasing values of the interaction parameter $\alpha$ lead to a sequence of curves in the phase diagram which monotonically approach the critical point; {\em ii)} the mean coupling $\bar J$ increases with bin size $\Delta t$, except for a small region of the parameter space in the case $\mu > \nu$; {\em iii)} $\bar h$ is not monotonic with $\Delta t$. In particular for $\alpha > \nu/2$ it decreases for $\Delta t$ large. Interestingly, the points which are inferred can correspond both to stable and metastable states. The latter occurs, for Hawkes processes, for large $\Delta t$ and $\alpha > \nu/2$.

Inference of financial data results in a trajectory similar to that of Hawkes processes with $\alpha>\nu/2$ (see Fig. \ref{fig:ResultCP}).

In all these cases, one can introduce a fictitious inverse temperature $\beta$, rescale the inferred couplings as
$(J_{ij},h_{i}) \rightarrow (\beta J_{ij},\beta h_{i})$, and analyze the corresponding statistical behavior. 
The fluctuations of observables as a function of $\beta$ provide an indication of the proximity of the inferred model to  a critical point. In figure \ref{fig:TD} we plot
the specific heat $c_V  = \beta^2 \, {\rm Var}[H(\vec s)]$
for various bin sizes in the case of financial data. For a given value of $\Delta t$, varying the inverse temperature $\beta$ corresponds to moving on the line passing through the origin and the inferred point (white line in Fig. \ref{fig:ResultCP}). If $\Delta t$ is in the region close to the critical point, for the reasons stated above, then fluctuations will be maximal for $ \beta \approx 1$. We remark, however, that such a notion of proximity to a critical point is only apparent. The distance from the critical point evaluated using $\beta$ is not invariant under reparametrization of the couplings: the number of distinguishable models in a given interval of 
temperature is not constant throughout the space of couplings.

\section{Discussion}

In summary, we have shown that the measure of distinguishable distribution in a parametric family of models is directly related to the susceptibility of the corresponding model in statistical mechanics. As  a consequence, this measure exhibit a singular concentration at critical points. One may speculate that, if experiments are designed (or data-set collected) in order to be maximally informative, they should return data which sample uniformly the space of distributions. This, as we have seen, corresponds to sampling a measure in parameter space which is sharply peaked at critical points. Hence, inference of data from maximally informative experiments (see \cite{Atkinson1992} for a survey) is likely to return parameters close to critical points with high probability. 

As stated in the introduction, critical points separate a region (or ``phase'') of weak interaction, where the different components behave in an essentially uncorrelated fashion, from a strongly interactive phase, where the knowledge of the microscopic variables in one part of the system fixes the state also in far away regions of the system. The critical point shares properties of both phases. It has the largest possible entropy consistent with system wide coherence. Often, system wide coherence is implicitly enforced by the fact that we construct data-sets with elements we believe to be mutually dependent or causally related. We would hardly analyze data-set with uncorrelated variables (e.g. the activity of a cell, planetary motion and fluctuations in financial markets). Therefore, criticality might not only come from the actual dynamics of the system, but it might be in the eyes of those who are trying to infer the underlying dynamical mechanisms.

Furthermore, if the inference depends on parameters which can be adjusted (such as $\Delta t$ above), then it is sensible to fix these parameters in a way which makes the determination of uncertainty about the model as small as possible. By Cramer-Rao bound, this again suggests that our inference should fall close to critical points. 

At the same time, concluding that a model is close to a critical point on the basis of a maximum of the specific heat in a plot like the one in Fig. \ref{fig:TD} can be misleading. Indeed the distance from the critical point should be measures in terms of the number of distinguishable models which the number of samples allow us to distinguish. It might be that even if the model is close to a critical point in the space of $\underline g$ (i.e. $|\underline g-\underline g_c|\ll 1$) there are many models between $\underline g$ and $\underline g_c$, which are closer to the critical point and which could, in principle, be distinguished on the basis of the data.

Even in the simple example presented here, there are some collective properties of the inferred states which turn out not to correspond to properties of the real system. The proximity to criticality is one such feature, since the original model (Hawkes process) does not have a proper phase transition. 
A further spurious feature is the fact that, in a region of parameters (large $\Delta t$ and $\alpha>\nu/2$) , the inferred model exhibits a double peaked distribution, but the empirical data is reproduced by the least probable maximum. This is the analog of metastability in physics, a phenomenon by which a system may be driven to attain a phase which is different from the one which would be stable in those circumstances. Metastable states usually decay in stable states, which would lead to the wrong expectation of a sharp transition in the system of which we're inferring the couplings. Actually, the distribution for the real system in this case does not have a second peak corresponding to that of the inferred model.

The increasing relevance of methods of statistical learning of high-dimensional models from data in a wide range of disciplines, makes it of utmost importance to understand which features of the inferred models are induced solely by the inference scheme and which ones reflect genuine features of the real system. In this respect, the understanding of collective behavior of models of statistical mechanics provides a valuable background. This is particularly true, in the presence of phase transitions of the associated statistical mechanics model, where the mapping between microscopic interaction and collective behavior is no longer single valued. 
The emphasis which the study of phase transitions and critical point phenomena has received in statistical physics, assumes a special relevance for inference, in the light of our findings. 




\begin{thebibliography}{99}



\bibitem{Braunstein2008} A.~Braunstein, {\em et al.} J. Stat. Mech. P12001 (2008).

\bibitem{Socolich2005}
  M.~Socolich, {\em et  al.} 
  Nature {\bf 437}, 512 (2005); M Weigt, {\em et al.}
  Proc. Natl. Acad. Sci. U.S.A. {\bf 106}, 67 (2009).

\bibitem{Schneidman2006}
  E.~Schneidman, {\em et  al.} 
  Nature {\bf 440}, 1007 (2006); 
  J.~Shlens, {\em et  al.} 
  J. Neurosci. {\bf 26}, 8254 (2006).
 
 \bibitem{Cocco2009}  
  S.~Cocco, S.~Leibler and R.~Monasson, Proc. Natl. Acad. Sci. U.S.A. {\bf 106}, 14058 (2009).

\bibitem{Lillo2008}
  F.~Lillo, E.~Moro, G.~Vaglica and R.~Mantegna,
  New J. Phys.Ê{\bf 10}, 043019 (2008);
  E.~Moro, {\em et  al.} 
  Physical Review E {\bf 80}, 066102 (2009);
  D.M.~ de Lachapelle and D.~Challet,
  New J. Phys. {\bf 12}, 075039 (2010).

\bibitem{Ackley1985}
  D.H.~Ackley, G.E.~Hinton and T.J.~Sejnowski,
  Cogn. Sci. {\bf 9}, 147 (1985).

\bibitem{Kappen1998}
  H. J.~Kappen and F.B.~Rodriguez,
  Neural Comput. {\bf 10}, 1137 (1998);
 T.~Tanaka,
 Phys. Rev. E {\bf 58}, 2302 (1998);
  V.~Sessak and R.~Monasson,
  J. Phys. A {\bf 42}, 055001 (2009);
  Y.~Roudi, J.~Tyrcha and J.~Hertz,
  Phys. Rev. E {\bf 79}, 051915 (2009);
  Y.~Roudi, E.~Aurell, J.A.~Hertz,
  Front. Comput. Neurosci. {\bf 3}, 22 (2009).
 

  
\bibitem{Tkacik2006}
  G.~Tkacik,  {\em et  al.} 
  qÐbio.NC/0611072 (2006).
  

\bibitem{Stephens2008}
G.J.~Stephens, T.~Mora, G.~Tkacik and W.~Bialek,
arXiv:0806.2694 [qÐbio.NC] (2008).  

 \bibitem{Mora2010}
 T.~Mora and W.~Bialek,
 arXiv:1012.2242 [qÐbio.QM] (2010).
 
\bibitem{Bak1987}
  P.~Bak, C.~Tang and K.~Wiesenfeld,
  Phys. Rev. Lett. {\bf 59}, 381 (1987).
 
\bibitem{Balasubramanian1997}
  V.~Balasubramanian,
  Neural Comput. {\bf 9}, 349 (1997);
  I.J.~Myung, V.~Balasubramanian and M.A.~Pitt,
  Proc. Natl. Acad. Sci. U.S.A. {\bf 97}, 11170 (2000).
  
\bibitem{Amari1985}
  S.I.~Amari,
  {\it Differential Geometrical Methods In Statistics}
  (Springer-Verlag, Berlin, 1985).

\bibitem{Ruppeiner1995}
G.~Ruppeiner,
Rev. Mod. Phys. {\bf 67}, 605 (1995).


\bibitem{Zanardi2007}
P.~Zanardi, P.~Giorda and M.~Cozzini,
Phys. Rev. Lett. {\bf 99}, 100603 (2007).

\bibitem{Hawkes1971a}
  A.G.~Hawkes,
  Biometrika {\bf 58}, 83 (1971);
  J. R. Statist. Soc. B {\bf 33}, 438 (1971).
 
\bibitem{Mezard2009}
  M.~Mezard and A.~Montanari,
  {\it Information, Physics, and Computation}
  (Oxford University Press, Oxford, 2009).
 
\bibitem{MDLfoot}
Close to critical points the model is very complex, in the sense of Minimal Description Length \cite{Balasubramanian1997,Rissanen1984}: introducing a penalization related to the complexity of the model leads
in these regions to macroscopically expensive models. Critical regions have a big descriptive power, as any small shift in the couplings allows to describe many other configurations.


\bibitem{Rissanen1984}
  J.~Rissanen,
  IEEE Trans. on Information Theory {\bf 30}, 629 (1984);

  J.~Rissanen,
  The Annals of Statistics {\bf14}, 1080 (1986);

  J.~Rissanen,
  IEEE Trans. on Information Theory {\bf 42}, 40 (1996).

\bibitem{Bowsher2002}
  C.G.~Bowsher,
  Economics Discussion Paper No. 2002-W22 (Nuffield College, Oxford, 2002);
  L.~Bauwens and N.~Hautsch,
   in {\it Handbook of Financial Time Series Econometrics},
   edited by T. A.~Andersen, {\em et al.} 
   (Springer-Verlag, Berlin, 2008).

\bibitem{Rieke1997}
  F.~Rieke, {\em et al.}  
  {\it Spikes: Exploring the Neural Code} (MIT Press, Cambridge, 1997);
  W.~Truccolo,  {\em et al.} 
  J. Neurophysiol. {\bf 93}, 1074 (2005).

 \bibitem{Borghesi2005}
  C.~Borghesi, M.~Marsili, S.~Miccich\`e,
  Phys. Rev. E {\bf 76}, 026104 (2005).


\bibitem{Roudi2009a}
  Y.~Roudi, S.~Nirenberg and P.E.~Latham,
  PLoS Comput. Biol. {\bf 5}, e1000380 (2009).

\bibitem{Epps1979}
  T.~W.~Epps,
 Journal of the American Statistical Association {\bf 74}, 291 (1979).
 
\bibitem{HWKfoot}
The properties of inferred couplings $(J,h)$ a for fully connected Hawkes process can be derived analytically in an approximate scheme (see lines in Fig. \ref{fig:ResultCP} (M.Marsili, I. Mastromatteo, in preparation).
 
 \bibitem{Atkinson1992}
   A.C.~Atkinson and A.N.~Donev,
   {\it Optimum experimental designs}
   (Claredon Press, New York, 1992).
 
%
%
%
%

\end{thebibliography}
\end{document}